\documentclass[fleqn,usenatbib]{mnras}

\usepackage[T1]{fontenc}

\DeclareRobustCommand{\VAN}[3]{#2}
\let\VANthebibliography\thebibliography
\def\thebibliography{\DeclareRobustCommand{\VAN}[3]{##3}\VANthebibliography}

\usepackage{graphicx}	
\usepackage{amsmath}	
\usepackage{amssymb}	
\usepackage{amsbsy,amsfonts}

\usepackage{psfrag}

\title[Eclipsing black hole in Sgr\,A*]{Modeling the shadow of Sgr\,A* through an eclipsing black hole}

\author[M.J. Santiba\~nez-Armenta et al.]{%
Milton Jair Santiba\~nez-Armenta,\thanks{E-mail:
msantibanez@astro.unam.mx (MJSA)}
Gustavo Magallanes-Guij\'on,\thanks{Email:
gmagallanes@astro.unam.mx (GMG)}
\newauthor
Sergio Mendoza\thanks{Email:
sergio@astro.unam.mx (SM)} \& 
Alejandro Cruz-Osorio\thanks{Email:
aosorio@astro.unam.mx (ACO)}
\\
Instituto de Astronom\'{\i}a, Universidad Nacional
                 Aut\'onoma de M\'exico, AP 70-264, Ciudad de M\'exico 04510,
                 M\'exico
}

\date{Accepted XXX. Received YYY; in original form ZZZ}

\pubyear{2024}

\begin{document}
\label{firstpage}
\pagerange{\pageref{firstpage}--\pageref{lastpage}}
\maketitle

\begin{abstract}
The Event Horizon Telescope (EHT) observations of Sgr\,A* resolved
the shadow image and emission ring-like structure,
which is associated to the photon ring of the supermassive black hole,
at the galactic centre, revealing a diameter of $51.8~\mu \text{as}$.
The ring-like structure is
consistent with that of a Kerr black hole. However, the source of the
high bright regions in the image and the time variability remain an
open question. Besides the plasma properties and emission models, the
spacetime geometry also holds an important role. We present an image
depicting the bright hot spots consistent with Sgr\,A*  observations
at a wavelength $ \lambda = 1.3\, {\rm mm}$. The image is the result
of an eclipsing Schwarzschild black hole situated along the line of
sight between the galactic centre and Earth.  The separation from both,
primary (Sgr\,A*) and secondary (eclipsing) black holes is $10233\, {\rm
AU}$. The central supermassive black hole located at the centre of the
galaxy has an observational  mass of $4.14\times10^6\,M_{\odot}$ and the
secondary eclipsing black hole has an inferred mass of $1035\,M_{\odot}$.

\end{abstract}





\section{Introduction}
\label{introduction}

The galactic centre, commonly referred as Sagittarius A* (Sgr\,A*),
has been extensively observed using various astronomical instruments
across different parts of the electromagnetic 
spectrum~\citep{Radio_observations,Adams_2021,1978ApJ...219L.101H}. Notably, 
Very Long Baseline Interferometry (VLBI) techniques have allowed for a 
detailed imaging and characterisation of this region. Using VLBI, it has
been possible to resolve a small region of the galactic centre using an
extended array of telescopes called the Event Horizon Telescope (EHT)
which produced a detailed image of the shadow of a supermassive black
hole surrounded by an accretion flow~\citep{EHT_SgrA_PaperI}. The image
clearly shows three bright hot spots around the 
emission ring-like structure, which is associated to the photon ring of black
hole.

In this work we report the shadow produced by the presence of a second
Schwarzschild black hole located between Sgr\,A* and Earth. The secondary
black hole with mass of $1035\,M_\odot$,  produces a lensing effect
on the light beams emitted at the accretion disc around the primary
supermassive black hole with mass of $4.14\times10^6\,M_{\odot}$ and
generates an eclipsed image of the Sgr\,A* shadow, similar to the one
produced by the~\citet{EHT_SgrA_PaperV}.

The first step in our modelling involves a general relativistic 
magnetohydrodynamics (GMHD) simulation of a Kerr black hole spacetime 
with dimensionless spin parameter $a_{\star}=15/16$ and accretion disc with 
constant specific angular momentum $ \ell=6.76 $, using the code \texttt{BHAC
v1.1}~\citep{Porth2017,Olivares2019}.  The rotating black hole
has a hydrodynamic equilibrium torus surrounding it which is perturbated
by a weak single-loop poloidal magnetic field  defined by its inner edge 
$r_\text{in}=20\, M$ and central radius $r_\text{c}=40 \, M$
~\citep{Fishbone76,Font02b,Shiokawa2012,Rezzolla_book:2013,Cruz2020}. 
The plasma is described by a
relativistic gas with an ideal equation of state~\citep[see
e.g.][]{Rezzolla_book:2013}, with adiabatic index $\Gamma=4/3$.

To compute the millimetre image of the accretion flow about
Sgr\,A* the radiative-transfer equation along null geodesics,
where the electromagnetic radiation propagates~\citep{Younsi2012,Younsi2020}, is
solved. We computed 500 snapshots covering 
$\approx 38\ {\rm h}$ of observations given the estimated mass
of Sgr\,A* of $4.14\times10^6\,\mathrm{M}_{\odot}$. The
image size is computed adopting a distance to Sgr\,A* of $8.127\,{\rm
kpc}$~\citep{EHT_SgrA_PaperI, EHT_SgrA_PaperV}.
The radiative transfer calculations require the use of a non-thermal 
energy  distribution for the modelled electrons~\citep{Davelaar2018,Fromm2021b,
Cruz2022,Roeder2023,Zhang2024}, which is a combination of a thermal  electron 
population and of a population with a power-law energy distribution inspired by 
particle-in-cell simulations~\citep[cf.][]{Ball2018a,Meringolo2023,Imbrogno2024}, 
with injection radius $r_{\rm inj}=10$ and the fraction of magnetic energy contributing 
to the heating of the radiating electrons $\varepsilon=0.5$. The electron temperature 
is computed using the R-$\beta$ model \citep{Moscibrodzka2009}.
To compute the mass accretion rate, we normalise the emission at a resolution of
$600\times600$ pixels, which corresponds to a field of view of $150\, \mu
{\rm as}$ in order to reproduce the observed flux of Sgr\,A* at 230\,GHz
of $\simeq 2.4\,{\rm Jy}$~\citep{EHT_SgrA_PaperI}. The resulting mass
accretion rates are $\dot{M}=[1.41,\, 1.37,\,  1.32, \, 1.29] \times
10^{-8}\, M_{\odot}\, \mathrm{yr}^{-1}$, respectively. The inclination
angle of the observation is assumed to be fixed at $0^{\circ},\,
30^{\circ},\, 45^{\circ},\, 90^{\circ}$, as deduced from the recent
observations of the EHT~\citep{EHT_SgrA_PaperV}.
Under these assumptions, only two of the hot spots in Sgr\,A* can be
reproduced, so far, the numerical GRMHD simulations considering a single black 
hole have not been capable of reproducing the observed third hot spot.

In the present work we construct a model in which a secondary eclipsing 
black hole with a mass \( \approx 1000\, \textrm{M}_\odot \) at a
distance of \( \approx 1200\, \textrm{AU} \) from the primary black hole
in Sgr\,A* is capable of bending light rays.
This secondary eclipsing black hole is contained in the line
of sight between the galactic centre and Earth and as shown in Figure~\ref{figure02} is capable of bending
the original shadow of Sgr\,A* in such a way as to reproduce the third
bright hot spot.

\section{Eclipsing black holes}
\label{methods}

\begin{figure}
\begin{center}
  \includegraphics[width=0.5\textwidth]{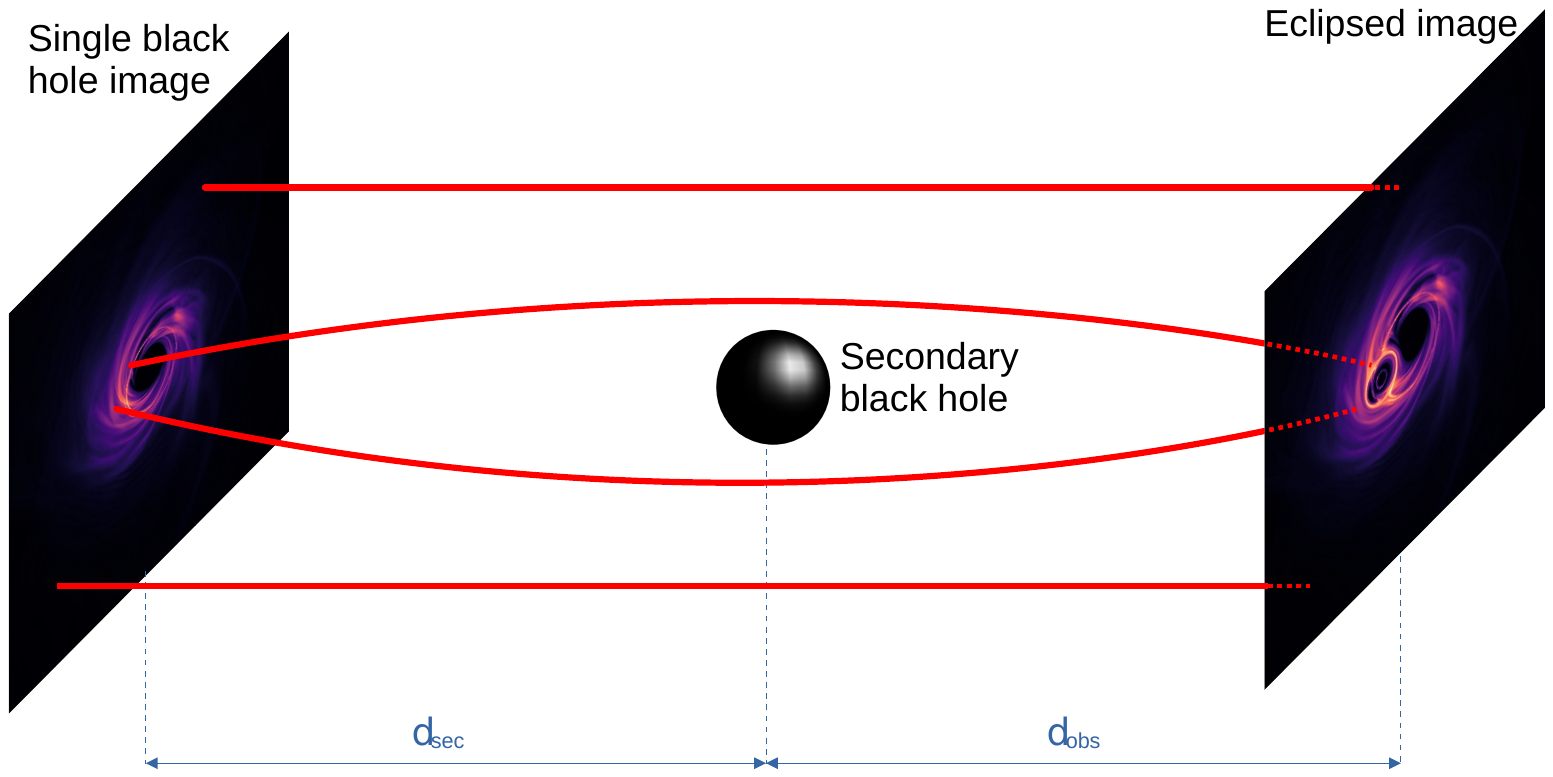}
 \end{center} 
  \caption{Artistic representation of a secondary Schwarzschild black
	hole with mass $M_\text{sec}$ eclipsing the shadow of Sgr\,A*. Red lines
represent light trajectories which are emitted from the single black hole image, lensed by  the secondary 
black hole and arriving to Earth as an eclipsed image. The distances from the secondary 
black hole to the image of Sgr\,A* and the one from the secondary black 
hole to the camera are represented by $d_{\rm sec}$ and $d_{\rm obs}$, respectively.   }
\label{figure02}
\end{figure}

The eclipsed image was generated by a secondary Schwarzschild black hole
using the recently developed numerical code \texttt{aztekas-shadows},
which evolves the geodesic equations using a 4th order Runge-Kutta
method~\citep[see e.g.][]{computational-methods-physicists}, solving the
schematic configuration shown in Figure~\ref{figure02}.  The algorithm
consists of emitting perpendicular light beams from a detector grid with
the same size and resolution as that of Sgr\,A*'s images of $600\times600$
pixels, which corresponds to \( 150 \mu\text{as}\) field of view, located
at a distance $d_{\mathrm{obs}}$ from the secondary black hole.  Those beams are
deflected by the secondary black hole and some of them would intersect the image
of Sgr\,A* located at a distance $d_\mathrm{sec}$ from the secondary black
hole, in the opposite direction of the grid.  The distances $d_\mathrm{obs}$
and $d_\text{sec}$ are greater than $2GM_{_\mathrm{sec}}/c^2\times10^4$, which
is $10^4$ times the gravitational radius $r_\text{g,\text{sec}}$ of secondary black hole,
where its gravitational influence is essentially negligible~\citep[see
e.g.][]{tesis_BOSS}. Therefore, the plane where the image of Sgr\,A* is located
and the detector are parallel to each other.  Every light beam which
intersects the image of Sgr\,A* was marked with the same corresponding
intensity on the detector grid.  The intensity was computed with the
original pixels of the image of Sgr\,A* using a bilinear interpolation.
The image obtained in the detector grid is thus the eclipsed shadow.
In other words, this final image is the shadow of the shadow of Sgr\,A*.

\begin{figure}
\begin{center}
\includegraphics[width=0.3\textwidth]{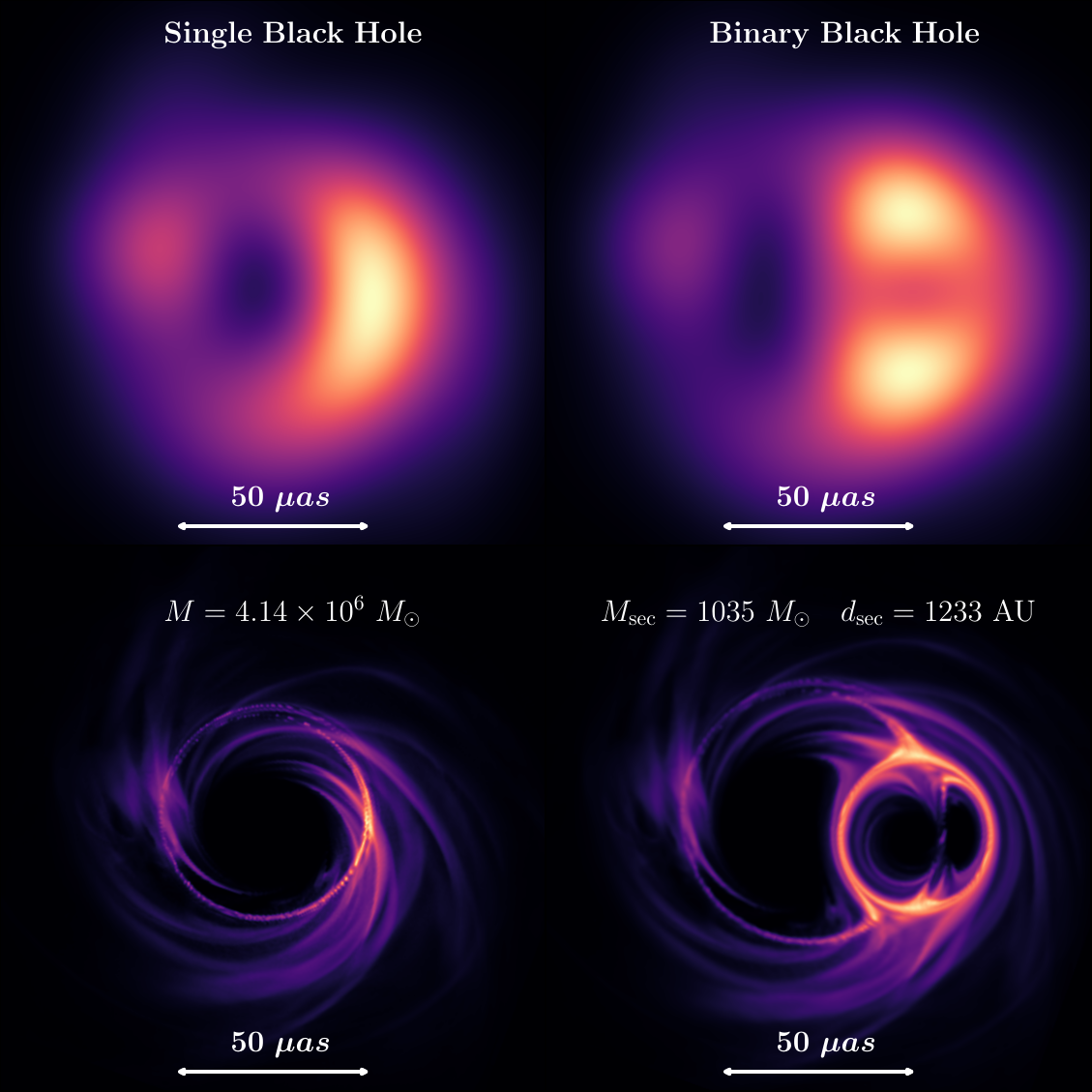}
\end{center}
\caption{ The left bottom panel shows the shadow of the primary Kerr black
hole in Sgr\,A* with $a_{\star}=15/16$, mass $M_{\rm BH}=4.14\times 10^{6}\ M_{\odot}$ 
and an inclination angle of $30^{\circ}$ with respect to the black hole spin axis.  
The right bottom panel is the  shadow of the left botton panel produced by an eclipsing 
Schwarzschild black hole. The top panels represent the blurring of their 
corresponding bottom images using a circular beam with radio of $20\, \mu as$. 
The mass ratio between supermassive and the secondary black hole is $q=2.5\times10^{-4}$, 
and the separation distance is $d_{\rm sec}=1233\,{\rm AU}$.}
\label{figure03}
\end{figure}

\begin{figure*}
\begin{center}
\includegraphics[width=0.7\textwidth]{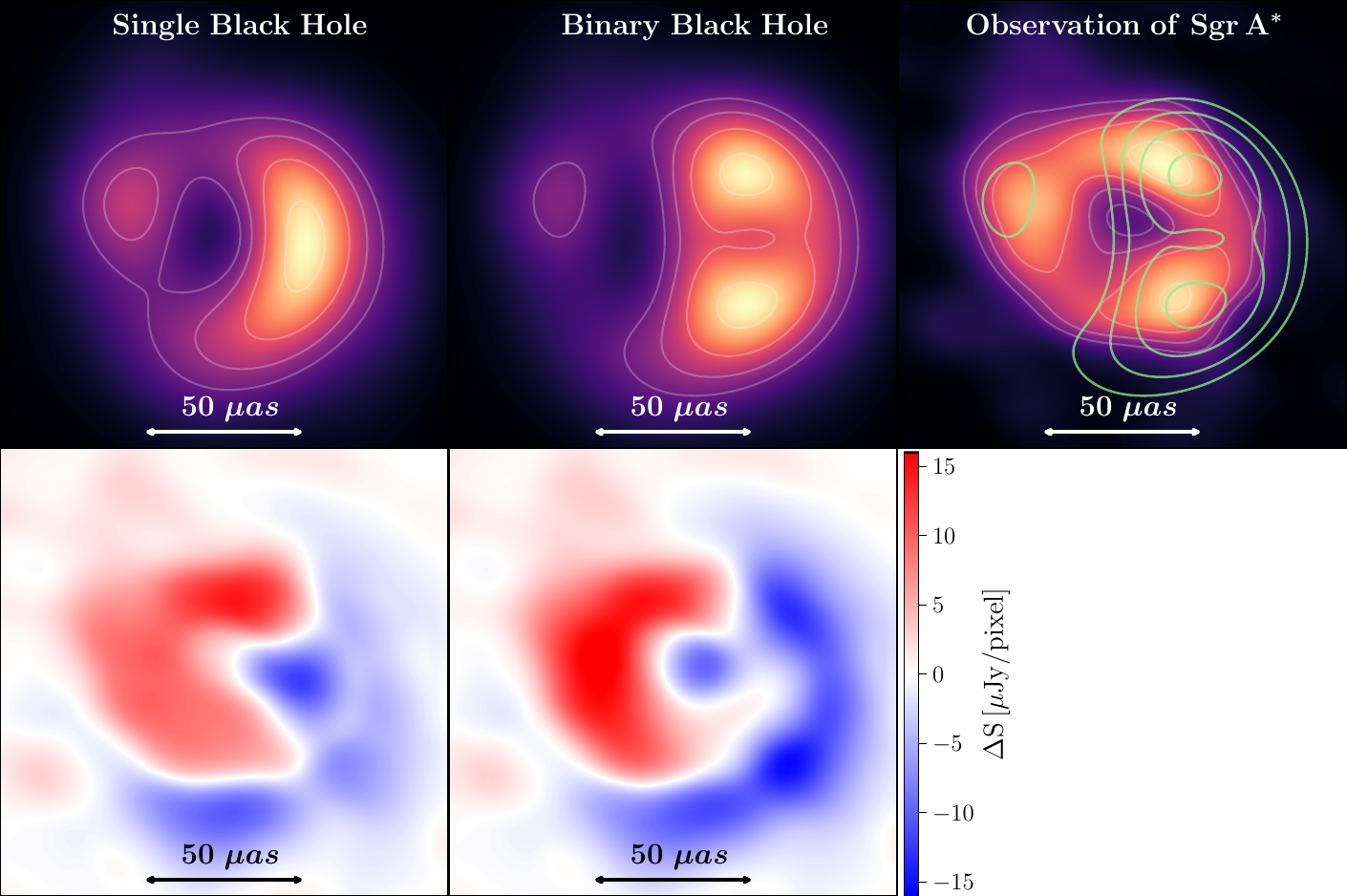}
\end{center}
\caption{The top left panel shows the synthetic image a single Kerr black hole. 
The top middle panel is the shadow of the top left panel produced by an eclipsing 
Schwarzschild black hole. The top right is the observed image of Sgr\,A* of the EHT 
colaboration~\citep[][]{EHT_SgrA_PaperI}, with green contours of the top middle 
image superimposed.
The synthetic images  were blurred using a circular beam of radio $20\, \mu as$ 
and the white lines represent constant flux contours with values of [$2.56,3.62,5.12,7.24$] 
$\times10^{-5}\mathrm{Jy}$. The mass of the Schwarzschild black hole is $M_{\rm BH}=
1035\ M_{\odot}$ so that the ratio between the primary to the secondary black 
hole is $q\approx2.5\times10^{-4}$, and the separation distance between them 
is $d_{\rm sec}=1233\,{\rm AU}$.
The bottom panels shown the relative errors between the observation of Sgr\,A*
and the synthetic images, $\Delta S= S_{\rm obs}-S_{\rm syn}$.}
\label{figure01}
\end{figure*}

We selected the best image that shows two clear and well-defined hotspots,
corresponding to a supermassive black hole as described above 
and an inclinations angle of $i=30^{\circ}$ between the spin axis and
Earth. Then we added a secondary Schwarzschild black hole with masses \(
M_\text{sec} \in \left[10^2M_{\odot},\, 10^5M_{\odot}\right] \), placed at
various distances $ d_\text{sec}\in[0.2\,AU,2000\,AU]$.  The distance \(
d_\text{obs} \) from the secondary black hole to Sgr\,A* image was always
taken with a fixed value of $r_{g,\text{sec}}\times10^5$ for every case.

We computed $\approx 100,000$ synthetic images of the eclipsing phenomenon,
by exploring different masses and locations of the secondary black hole. 
In particular we explored coplanar circular orbits by varying the radial direction $r\in[0 \, M, 15\,M]$ and 
azimuthal position $\varphi\in[0^\circ,360^\circ]$ around the primary black hole.
In order to significantly speed up the ray-tracing process to obtain 
the required images, the program \texttt{aztekas-shadows} required a 
parallelisation procedure in its development. Using  \(\approx 16,000 \) GPU  
units of an NVIDIA card with Compute Unified Device Architecture (CUDA), 
the computational time of the numerical simulations was reduced by five 
orders of magnitude. 

The distances $d_\mathrm{sec}$ and $d_\mathrm{obs}$ on the
eclipsed images generated by \texttt{aztekas-shadows} are much
greater than the gravitation radius of the secondary black hole
($d_\mathrm{sec},d_\mathrm{obs}>r_\text{g,\text{sec}} \times10^3$).  While,
the mass of the secondary black hole $M_{_\mathrm{sec}}\lesssim 2.5
M \times 10^{-2}$ is two orders of magnitude smaller than the central 
supermassive black hole. 
Since the analysed light beams are detected practically in a perpendicular
direction to the detector, their deflection angles are very small.  As a
consequence, a length scale $l_i$ on the image's grid will be modified as
$l_f$ on the image of Sgr\,A*  so that  $ l_i-l_f\propto d_\text{sec}q$.
In this way, we can reproduce similar images as long as the length-factor
product $d_\mathrm{sec}\,q$ is constant. 
In Figure~\ref{figure03}, we present the accretion flow around a single black hole, 
producing a synthetic image with two distinct brightness regions (top-left). 
In contrast, the eclipsed image—resulting from the presence of a secondary 
black hole and representing the best-fitting model—exhibits three hot spots (top-middle), 
aligning well with the EHT observations (top-right). The bottom panels display 
the relative errors between the synthetic images and the observational data, 
highlighting the differences between the models.


 \section{Summary and conclusions}
\label{discussion}

  We explored different configuration positions and masses of the
secondary eclipsing black hole following the constraints imposed
by~\citet{will2023}.  The first image of the shadow of Sgr\,A* was
constructed numerically by considering only the supermassive 
black hole at the galactic centre and the emission from the magnetised 
accretion disc. We assumed that the photons travel in straight paths 
at the distance of the secondary black hole, so that the light rays are bent 
when they pass near the secondary black hole. We found that the best 
suitable candidate to achieve the production of the third hot spot can be 
done by a secondary
Schwarzschild black hole of mass \( M_\text{sec} = 1035\, \text{M}_\odot \) at a
distance of \( d_\text{sec} = 10233 \, \text{AU} \) from the galactic centre as
shown in Figure~\ref{figure01}.  Smaller secondary black holes can also
account for this third hot spot if they follow the simple analytical
approximation given by \( d_\text{sec}q \approx 0.308 \, \text{AU}\),
where \( q = M_\text{sec} / M \), \( M_\text{sec} \) is the mass of
the secondary black hole and \( d_\text{sec} \) the separation distance
between the black holes.


The synthetic image obtained from a lensed shadow image due to the 
presence of a secondary intermediate black hole in the galactic centre 
has an overall good agreement with the reconstructed average image 
from the EHT observations of Sgr\,A*.  We found that the binary system 
with separation of $10233\, {\rm AU}$ and a secondary black hole with 
mass of $1035\,M_{\odot}$ produces a lensed image with three hot spots.
Figure~\ref{figure01} shows a direct image comparison between the observed
image of Sag\,A* with the numerical constructed images with and without a
secondary black hole, showing that the image with a secondary black hole is
a better model.
Our results are also in agreement with the recent theoretical constraints 
of the distance and mass of the secondary compact object.  The study 
performed by \citet{will2023} using the Newtonian three-body problem 
considering the binary black hole system and the S2 star orbit found two 
constriction regimes for the S2 orbit not to be disrupted. The first one 
is when the secondary black hole is farther away than the S2 star orbit 
at $\sim 1020\, {\rm AU}$  finding that distances \( d_\text{sec} \) in the 
range of $d_\mathrm{sec}\in[1000\,\mathrm{AU},4000\,\mathrm{AU}]$ 
and black hole masses $M_{_\mathrm{sec}}\in[10^3\,\mathrm{M}_{\odot},
10^5\,\mathrm{M}_{\odot}]$. The second region is inside the S2 orbit and
excludes the masses for the secondary black hole in the same range, 
 $M_{_\mathrm{sec}}\in[10^3\,\mathrm{M}_{\odot},10^5\,\mathrm{M}_{\odot}]$.
These limits ensure that the orbit of S2 is not affected by quadrupolar 
perturbations and the amplitude of gravitational waves generated by the 
binary black holes is negligible. This means that in the search parameter 
space for the simulations ($ d_\text{sec} > 1020, {\rm AU} $), a smaller value 
of $d_\mathrm{sec}$ requires also a smaller value of  $M_{_\mathrm{sec}}$. 
Our lensed synthetic image from the binary system satisfies these constraints. 

Since the angular diameter \( d_{sh} \) of Sgr~A* is given by
\citep{EHT_SgrA_PaperI}: \(d_{sh} \approx 10 GM / c^2 D \approx 48.7 \mu
\text{as}\) and the angular diameter of the eclipsing secondary black
hole turns out to be \( 0.013 \mu \text{as} \), the ratio between the
secondary black hole and Sgr A* is \( \approx 2.7 \times 10^{-4} \).
This result clearly shows that the eclipse is partial and localised in
order to get the required third hot spot.

Finally, in our approach, the eclipsing binary black hole could be
a fly-by object or an orbiting one about the galactic centre.  In the
latter case for a Keplerian circular orbit, with a radius $ \approx
10^4\, {\rm AU}$ around the galactic centre, it would have an orbiting
period of \( \approx 26\, \text{yr} \).  In general terms, 
we expect the third hot spot to be absent in the next release of Sgr A*
observations if the secondary black hole is in orbit about the supermassive black hole, Sgr A*.

The results presented here
highlight the need for further research and enhancements, particularly
in modelling that incorporates the self-consistent dynamics of binary
black holes surrounded by an accretion flow, and the analysis of the
time variability of the light curve of Sgr\,A*.

\section*{Acknowledgments}
\label{acknowledgements}

This work was supported by PAPIIT DGAPA-UNAM projects (IN110522,
IN118325, IA103725).  ACO, GMG, SM, MJSA acknowledge economic support from
``Secretaría de Ciencia, Humanidades, Tecnolog\'{\i}a e Innovaci\'on''
(Secihti, M\'exico), numbers: 257435, 378460, 26344, 751147.  GMG and
MJSA acknowledge economic support from CGEP-UNAM for a research visit
at Goethe Universit\"{a}t, Frankfurt.  ACO gratefully acknowledges
``Ciencia Básica y de Frontera 2023-2024'' program of Secihti
(CBF2023-2024-1102) and the European Horizon Europe staff exchange (SE)
programme HORIZON-MSCA2021-SE-01 Grant No. NewFunFiCO-101086251.


%
%


\section*{Data Availability}
The data that support the plots within this paper
and other findings of this study are available on:
\url{https://archive.org/details/sagAbinaryblackhole}.  The public
released version of the GRMHD code \texttt{BHAC} can be found at
\href{https://bhac.science}{https://bhac.science}.  The public version
or the code  \texttt{aztekas-shadows} will be released soon.



\bibliographystyle{mnras}
\bibliography{saga}

\begin{thebibliography}{}
\makeatletter
\relax
\def\mn@urlcharsother{\let\do\@makeother \do\$\do\&\do\#\do\^\do\_\do\%\do\~}
\def\mn@doi{\begingroup\mn@urlcharsother \@ifnextchar [ {\mn@doi@}
  {\mn@doi@[]}}
\def\mn@doi@[#1]#2{\def\@tempa{#1}\ifx\@tempa\@empty \href
  {http://dx.doi.org/#2} {doi:#2}\else \href {http://dx.doi.org/#2} {#1}\fi
  \endgroup}
\def\mn@eprint#1#2{\mn@eprint@#1:#2::\@nil}
\def\mn@eprint@arXiv#1{\href {http://arxiv.org/abs/#1} {{\tt arXiv:#1}}}
\def\mn@eprint@dblp#1{\href {http://dblp.uni-trier.de/rec/bibtex/#1.xml}
  {dblp:#1}}
\def\mn@eprint@#1:#2:#3:#4\@nil{\def\@tempa {#1}\def\@tempb {#2}\def\@tempc
  {#3}\ifx \@tempc \@empty \let \@tempc \@tempb \let \@tempb \@tempa \fi \ifx
  \@tempb \@empty \def\@tempb {arXiv}\fi \@ifundefined
  {mn@eprint@\@tempb}{\@tempb:\@tempc}{\expandafter \expandafter \csname
  mn@eprint@\@tempb\endcsname \expandafter{\@tempc}}}

\bibitem[\protect\citeauthoryear{Adams et~al.,}{Adams
  et~al.}{2021}]{Adams_2021}
Adams C.~B.,  et~al., 2021, \mn@doi [The Astrophysical Journal]
  {10.3847/1538-4357/abf926}, 913, 115

\bibitem[\protect\citeauthoryear{{Ball}, {Sironi}  \& {{\"O}zel}}{{Ball}
  et~al.}{2018}]{Ball2018a}
{Ball} D.,  {Sironi} L.,   {{\"O}zel} F.,  2018, \mn@doi [Astrophys. J.]
  {10.3847/1538-4357/aac820}, \href
  {https://ui.adsabs.harvard.edu/abs/2018ApJ...862...80B} {862, 80}

\bibitem[\protect\citeauthoryear{{Cruz-Osorio}, {Gimeno-Soler}  \&
  {Font}}{{Cruz-Osorio} et~al.}{2020}]{Cruz2020}
{Cruz-Osorio} A.,  {Gimeno-Soler} S.,   {Font} J.~A.,  2020, \mn@doi [Mon. Not.
  R. Astron. Soc.] {10.1093/mnras/staa216}, \href
  {https://ui.adsabs.harvard.edu/abs/2020MNRAS.492.5730C} {492, 5730}

\bibitem[\protect\citeauthoryear{{Cruz-Osorio} et~al.,}{{Cruz-Osorio}
  et~al.}{2022}]{Cruz2022}
{Cruz-Osorio} A.,  et~al., 2022, \mn@doi [Nature Astronomy]
  {10.1038/s41550-021-01506-w}, \href
  {https://ui.adsabs.harvard.edu/abs/2022NatAs...6..103C} {6, 103}

\bibitem[\protect\citeauthoryear{{Davelaar}, {Mo{\'s}cibrodzka}, {Bronzwaer}
  \& {Falcke}}{{Davelaar} et~al.}{2018}]{Davelaar2018}
{Davelaar} J.,  {Mo{\'s}cibrodzka} M.,  {Bronzwaer} T.,   {Falcke} H.,  2018,
  \mn@doi [Astron. Astrophys.] {10.1051/0004-6361/201732025}, \href
  {http://adsabs.harvard.edu/abs/2018A%26A...612A..34D} {612, A34}

\bibitem[\protect\citeauthoryear{{Event Horizon Telescope Collaboration}
  et~al.}{{Event Horizon Telescope Collaboration}
  et~al.}{2022a}]{EHT_SgrA_PaperI}
{Event Horizon Telescope Collaboration} et~al., 2022a, \mn@doi [Astrophys. J.
  Lett.] {10.3847/2041-8213/ac6674}, \href
  {https://ui.adsabs.harvard.edu/abs/2022ApJ...930L..12A} {930, L12}

\bibitem[\protect\citeauthoryear{{Event Horizon Telescope Collaboration}
  et~al.}{{Event Horizon Telescope Collaboration}
  et~al.}{2022b}]{EHT_SgrA_PaperV}
{Event Horizon Telescope Collaboration} et~al., 2022b, \mn@doi [Astrophys. J.
  Lett.] {10.3847/2041-8213/ac6672}, \href
  {https://ui.adsabs.harvard.edu/abs/2022ApJ...930L..16A} {930, L16}

\bibitem[\protect\citeauthoryear{{Fishbone} \& {Moncrief}}{{Fishbone} \&
  {Moncrief}}{1976}]{Fishbone76}
{Fishbone} L.~G.,  {Moncrief} V.,  1976, Astrophys. J., \href
  {http://adsabs.harvard.edu/abs/1976ApJ...207..962F} {207, 962}

\bibitem[\protect\citeauthoryear{Font \& Daigne}{Font \&
  Daigne}{2002}]{Font02b}
Font J.~A.,  Daigne F.,  2002, Astrophys. J, 581, L23

\bibitem[\protect\citeauthoryear{{Fromm} et~al.,}{{Fromm}
  et~al.}{2022}]{Fromm2021b}
{Fromm} C.~M.,  et~al., 2022, \mn@doi [Astron. Astrophys.]
  {10.1051/0004-6361/202142295}, \href
  {https://ui.adsabs.harvard.edu/abs/2022A&A...660A.107F} {660, A107}

\bibitem[\protect\citeauthoryear{{Gallego-Calvente, A. T.}
  et~al.,}{{Gallego-Calvente, A. T.} et~al.}{2021}]{Radio_observations}
{Gallego-Calvente, A. T.} et~al., 2021, \mn@doi [A\&A]
  {10.1051/0004-6361/202039172}, 647, A110

\bibitem[\protect\citeauthoryear{{Hildebrand}, {Whitcomb}, {Winston},
  {Stiening}, {Harper}  \& {Moseley}}{{Hildebrand}
  et~al.}{1978}]{1978ApJ...219L.101H}
{Hildebrand} R.~H.,  {Whitcomb} S.~E.,  {Winston} R.,  {Stiening} R.~F.,
  {Harper} D.~A.,   {Moseley} S.~H.,  1978, \mn@doi [Astrophys. J. Lett.]
  {10.1086/182616}, \href
  {https://ui.adsabs.harvard.edu/abs/1978ApJ...219L.101H} {219, L101}

\bibitem[\protect\citeauthoryear{{Imbrogno}, {Meringolo}, {Servidio},
  {Cruz-Osorio}, {Cerutti}  \& {Pegoraro}}{{Imbrogno}
  et~al.}{2024}]{Imbrogno2024}
{Imbrogno} M.,  {Meringolo} C.,  {Servidio} S.,  {Cruz-Osorio} A.,  {Cerutti}
  B.,   {Pegoraro} F.,  2024, \mn@doi [Astrophys. J. Lett.]
  {10.3847/2041-8213/ad6b9d}, \href
  {https://ui.adsabs.harvard.edu/abs/2024ApJ...972L...5I} {972, L5}

\bibitem[\protect\citeauthoryear{{Meringolo}, {Cruz-Osorio}, {Rezzolla}  \&
  {Servidio}}{{Meringolo} et~al.}{2023}]{Meringolo2023}
{Meringolo} C.,  {Cruz-Osorio} A.,  {Rezzolla} L.,   {Servidio} S.,  2023,
  \mn@doi [Astrophys. J.] {10.3847/1538-4357/acaefe}, \href
  {https://ui.adsabs.harvard.edu/abs/2023ApJ...944..122M} {944, 122}

\bibitem[\protect\citeauthoryear{{Mo{\'s}cibrodzka}, {Gammie}, {Dolence},
  {Shiokawa}  \& {Leung}}{{Mo{\'s}cibrodzka} et~al.}{2009}]{Moscibrodzka2009}
{Mo{\'s}cibrodzka} M.,  {Gammie} C.~F.,  {Dolence} J.~C.,  {Shiokawa} H.,
  {Leung} P.~K.,  2009, \mn@doi [Astrophys. J.] {10.1088/0004-637X/706/1/497},
  \href {http://adsabs.harvard.edu/abs/2009ApJ...706..497M} {706, 497}

\bibitem[\protect\citeauthoryear{Olivares, Porth, Davelaar, Most, Fromm,
  Mizuno, Younsi  \& Rezzolla}{Olivares et~al.}{2019}]{Olivares2019}
Olivares H.,  Porth O.,  Davelaar J.,  Most E.~R.,  Fromm C.~M.,  Mizuno Y.,
  Younsi Z.,   Rezzolla L.,  2019, \mn@doi [Astron. Astrophys.]
  {10.1051/0004-6361/201935559}, 629, A61

\bibitem[\protect\citeauthoryear{{Porth}, {Olivares}, {Mizuno}, {Younsi},
  {Rezzolla}, {Moscibrodzka}, {Falcke}  \& {Kramer}}{{Porth}
  et~al.}{2017}]{Porth2017}
{Porth} O.,  {Olivares} H.,  {Mizuno} Y.,  {Younsi} Z.,  {Rezzolla} L.,
  {Moscibrodzka} M.,  {Falcke} H.,   {Kramer} M.,  2017, \mn@doi [Computational
  Astrophysics and Cosmology] {10.1186/s40668-017-0020-2}, 4, 1

\bibitem[\protect\citeauthoryear{{Rezzolla} \& {Zanotti}}{{Rezzolla} \&
  {Zanotti}}{2013}]{Rezzolla_book:2013}
{Rezzolla} L.,  {Zanotti} O.,  2013, Relativistic Hydrodynamics.
Oxford University Press, Oxford, UK,
  \mn@doi{10.1093/acprof:oso/9780198528906.001.0001}

\bibitem[\protect\citeauthoryear{{R{\"o}der}, {Cruz-Osorio}, {Fromm}, {Mizuno},
  {Younsi}  \& {Rezzolla}}{{R{\"o}der} et~al.}{2023}]{Roeder2023}
{R{\"o}der} J.,  {Cruz-Osorio} A.,  {Fromm} C.~M.,  {Mizuno} Y.,  {Younsi} Z.,
   {Rezzolla} L.,  2023, \mn@doi [Astron. Astrophys.]
  {10.1051/0004-6361/202244866}, \href
  {https://ui.adsabs.harvard.edu/abs/2023A&A...671A.143R} {671, A143}

\bibitem[\protect\citeauthoryear{{Shiokawa}, {Dolence}, {Gammie}  \&
  {Noble}}{{Shiokawa} et~al.}{2012}]{Shiokawa2012}
{Shiokawa} H.,  {Dolence} J.~C.,  {Gammie} C.~F.,   {Noble} S.~C.,  2012,
  \mn@doi [Astrophys. J.] {10.1088/0004-637X/744/2/187}, \href
  {http://adsabs.harvard.edu/abs/2012ApJ...744..187S} {744, 187}

\bibitem[\protect\citeauthoryear{Sirca \& Horvat}{Sirca \&
  Horvat}{2012}]{computational-methods-physicists}
Sirca S.,  Horvat M.,  2012, Computational Methods for Physicists: Compendium
  for Students.
Graduate Texts in Physics, Springer Berlin Heidelberg, Germany, \url
  {https://books.google.com.mx/books?id=hIPip3UxX3gC}

\bibitem[\protect\citeauthoryear{{Will}, {Naoz}, {Hees}, {Tucker}, {Zhang},
  {Do}  \& {Ghez}}{{Will} et~al.}{2023}]{will2023}
{Will} C.~M.,  {Naoz} S.,  {Hees} A.,  {Tucker} A.,  {Zhang} E.,  {Do} T.,
  {Ghez} A.,  2023, \mn@doi [arXiv e-prints] {10.48550/arXiv.2307.16646}, \href
  {https://ui.adsabs.harvard.edu/abs/2023arXiv230716646W} {p. arXiv:2307.16646}

\bibitem[\protect\citeauthoryear{Younsi}{Younsi}{2014}]{tesis_BOSS}
Younsi Z.,  2014, PhD thesis

\bibitem[\protect\citeauthoryear{{Younsi}, {Wu}  \& {Fuerst}}{{Younsi}
  et~al.}{2012}]{Younsi2012}
{Younsi} Z.,  {Wu} K.,   {Fuerst} S.~V.,  2012, \mn@doi [Astron. Astrophys.]
  {10.1051/0004-6361/201219599}, \href
  {http://adsabs.harvard.edu/abs/2012A%26A...545A..13Y} {545, A13}

\bibitem[\protect\citeauthoryear{{Younsi}, {Porth}, {Mizuno}, {Fromm}  \&
  {Olivares}}{{Younsi} et~al.}{2020}]{Younsi2020}
{Younsi} Z.,  {Porth} O.,  {Mizuno} Y.,  {Fromm} C.~M.,   {Olivares} H.,  2020,
  in {Asada} K.,  {de Gouveia Dal Pino} E.,  {Giroletti} M.,  {Nagai} H.,
  {Nemmen} R.,  eds, ~ Vol. 342, Perseus in Sicily: From Black Hole to Cluster
  Outskirts. pp 9--12 (\mn@eprint {arXiv} {1907.09196}),
  \mn@doi{10.1017/S1743921318007263}

\bibitem[\protect\citeauthoryear{{Zhang}, {Mizuno}, {Fromm}, {Younsi}  \&
  {Cruz-Osorio}}{{Zhang} et~al.}{2024}]{Zhang2024}
{Zhang} M.,  {Mizuno} Y.,  {Fromm} C.~M.,  {Younsi} Z.,   {Cruz-Osorio} A.,
  2024, \mn@doi [Astron. Astrophys.] {10.1051/0004-6361/202449497}, \href
  {https://ui.adsabs.harvard.edu/abs/2024A&A...687A..88Z} {687, A88}

\makeatother
\end{thebibliography}


\bsp	
\label{lastpage}
\end{document}